\documentstyle[prl,aps]{revtex}
\twocolumn

\begin{document}
\author{Jian Qi Shen $^{1,2}$ \footnote{E-mail address: jqshen@coer.zju.edu.cn, jqshencn@yahoo.com.cn} and Fei Zhuang $^{3}$}
\address{$^{1}$  Centre for Optical
and Electromagnetic Research, Joint Research Centre of Photonics
of the Royal Institute of Technology (Sweden) and Zhejiang
University, Zhejiang University,
Hangzhou Yuquan 310027, People's Republic of China\\
$^{2}$ Zhejiang Institute of Modern Physics and Department of
Physics, Zhejiang University, \\Hangzhou 310027, People's Republic
of China\\
$^{3}$ Department of Physics, Hangzhou Teacher's College, Hangzhou
310012,  People's Republic of China}
\date{\today}
\title{Propagation of atomic matter waves inside an atom wave guide}
\maketitle

\begin{abstract}
The phenomenological description of propagation of atomic matter
waves inside a curved atom wave guide is presented based on the
effective action principle. The evolutions in both temporal and
spatial domains for the atomic matter wave in the presence of
guiding potential fields (classical and quantized) are considered.
The coherent control of three-level atomic matter waves in a wave
guide by an external controlling light is briefly discussed. The
concepts of {\it atomic matter-wave bandgap structure} in a
spatially periodic guiding field ({\it e.g.}, the interior
potential of carbon peapod which can trap both atoms and light)
and {\it optical lattice bandgap medium} are suggested.
\\ \\
\textbf{PACS.} 42.50.Ct Quantum description of interaction of
light and matter - 03.65.Sq Semiclassical theories and
applications - 03.75.Be Atom and neutron optics
\end{abstract}
\pacs{}

\section{Introduction}
During the last few years, with the development of matter-wave
interferometry and atom-manipulation technologies ({\it e.g.},
guiding, cooling and trapping atoms), many rapid progresses have
been made in the atom optics (atomic matter-wave
optics)\cite{Badurek,Chu,Storey,Metcalf,Balykin}. The techniques
(including the atom optical devices) utilized for manipulating
atoms are as follows: magnetic confinement inside both hollow
glass tubes\cite{Noh} and current carrying wires\cite{Reichel1},
permanent micromagnets\cite{Ven}, light-induced force
trapping\cite{Gri} and microfabricated-structure method\cite{Bir}.
The controllable manipulation of cold neutral atoms in field
potentials created by substrate mounted microstructures has been
advancing at an increasing pace recently. Many matter-wave optical
devices such as tightly confining traps and
guides\cite{Reichel,Folman}, beam splitters\cite{Cassettari} and
mirrors\cite{HindsHughes} have been realized on such atom chips.
More recently, Kr\"{u}ger {\it et al.} suggested a scheme of
three-dimensional trap formed by modulating a magnetic guide using
electrostatic fields, and observed atoms trapped in a string of up
to six individual such traps\cite{Kruger}; Luo {\it et al.}
presented an omnidirectional matter wave guide (atom fibre) on an
atom chip, and demonstrated the guiding of thermal atoms around
more than two complete turns along a spiral shaped 25 mm long
curved path (curve radii down to 200 $\mu$m) at various
atom-surface distances (35-450 $\mu$m)\cite{Luo}. As for the
theoretical study of atom trapping and manipulating with
microfabricated structure (atom chips), J\"{a}\"{a}skel\"{a}inen
{\it et al.} considered many related topics, including the
localization in splitting of matter waves, adiabatic propagation
and reflection of atoms in potential structures, self-imaging in
atom wave guides, multimode interferometer for guided matter waves
as well as quantum-state measurement through ballistic expansion
of matter
waves\cite{Stenholm1,Stenholm2,Stenholm3,Folman5,Rohwedder,Stenholm4}.
Since most of these investigations concentrated their attention
primarily on the subjects of the matter-wave propagation in the
{\it straight} atomic-wave guides, but did not consider the topics
inside the curved (say, spiral shaped) guides\cite{Luo}, we
proposed a formulation to deal with the wave propagation problem
in the curved guides in Ref.\cite{Shen2}, which is a
phenomenological description of time evolution of atomic matter
waves inside a noncoplanar atom fibre (atomic-wave guide). In this
paper, we first establish a theoretical base for such a
phenomenological description, and then treat the evolutional
behavior of neutral atomic matter wave in both classical and
quantized potential fields. We discuss the evolutions of guided
atomic matter wave in both spatial and temporal domains. As one of
the most remarkable results of this treatment, the {\it atomic
matter-wave bandgap structure} in a cavity field and a spatially
periodic guiding field is considered. Here the concept of {\it
atomic bandgap medium} (optical lattice bandgap medium) is
analogous to the photonic bandgap material (photonic
crystal)\cite{Yablonovitch,John} and phononic crystal. Finally, we
suggest and briefly discuss a scheme of the controllable
manipulation of three-level atomic matter waves in a wave guide by
an external controlling light.

\section{Phenomenological description of propagation of atomic matter waves inside a spiral shaped atom wave guide}

In this section, based on the effective action principle, we will
suggest a phenomenological description of propagation of atomic
matter waves inside a spiral shaped atom wave guide (atom fibre).
It was verified in Ref.\cite{Shen2} that the effective interaction
that causes the curved-path motion of matter wave in the atom
fibre is a ``magnetic''-type (topological, global) interaction,
which has a vanishing action\cite{ShenAnn}. So, we should
construct such a vanishing action (or Lagrangian). In general, the
propagation of matter wave inside the guide contains the
transverse and longitudinal motions. The atomic motional state in
the transverse direction is confined (trapped) by a side-guide
potential ({\it e.g.}, magnetic-wire guide potential that is the
result of the homogeneous external bias field). Thus, atoms can
only be guided along the third unconfined (longitudinal)
direction. First we assume that the guiding potential allows us to
separate the atomic matter wavefunction, {\it i.e.}, in the wave
guide it can be rewritten as the product of the transverse state
and the longitudinal state. The transverse modes can be chosen to
be the standing wavefunctions or harmonic oscillator
eigenfunctions\cite{Stenholm2}, which we will not consider further
here. Instead, we are concerned with the atomic motion (wave
propagation) in the one-dimensional longitudinal direction. Thus,
the vanishing Lagrangian that characterizes the phenomenological
(effective) interaction between the guided matter wave and the
guiding potential fields can be written as $ L=\eta \hbar(kv-{\bf
k}\cdot{\bf v})=\eta \hbar \dot{\bf k}\cdot {\bf r}$, where dot
denotes the time derivative, and $v$ and $k$ the atomic velocity
and the propagation constant (wave vector), respectively, along
the guide path (one-dimensional longitudinal propagation). ${\bf
v}$ and $v$ are defined as ${\bf v}={\rm d}{\bf r}/{\rm d}t$ and
$v={\rm d}s/{\rm d}t$, where $s$ is the guide path length. Since
the direction of ${\bf k}$ is the same as that of the
one-dimensional longitudinal guide path, then one gets the
relation $ks={\bf k}\cdot{\bf r}$. For a matter wave, the
relationship between the wave vector and the particle velocity is
${\bf k}=\mu {\bf v}/\hbar$, where $\mu$ is the mass of the atom.
Thus one has $\mu{\bf v}\cdot{\bf k}/\left(\hbar k^{2}\right)=1$.
So, the phenomenological Lagrangian can be rewritten as $L=\eta
\hbar \left(\mu/\hbar k^{2}\right)\left({\bf v}\cdot{\bf
k}\right)\dot{\bf k}\cdot {\bf r}$. It is assumed that the modulus
of the wave vector ${\bf k}$ does not alter much in the evolution
process inside the curved atom wave guide, {\it i.e.}, $k^{2}$ can
be considered a constant number. Thus, we can have ${\bf
v}\cdot\dot{\bf k}=0$. If we add a vanishing term $-\eta \hbar
\left(\mu/\hbar k^{2}\right)({\bf v}\cdot\dot{\bf k}){\bf
r}\cdot{\bf k}$ to $L$, then the new Lagrangian is
\begin{equation}
L=\eta \mu{\bf v}\cdot\left[{\bf r}\times \left(\frac{{\bf
k}\times \dot{\bf k}}{k^{2}}\right)\right].
\end{equation}
Apparently, such a Lagrangian has a form $\mu{\bf v}\cdot{\bf A}$.
Here the three-dimensional effective vector potential is ${\bf
A}=\eta {\bf r}\times ({{\bf k}\times \dot{\bf k}})/{k^{2}}$, and
in consequence, the phenomenological (effective) field strength is
${\bf B}_{\rm eff}=\nabla\times{\bf A}=-2\eta ({{\bf k}\times
\dot{\bf k}})/{k^{2}}$. Further calculation shows that the
phenomenological Hamiltonian is $H=\left({\bf P}-\mu{\bf
A}\right)^{2}/2\mu$ with the canonical momentum being ${\bf
P}=\mu{\bf v}+\mu{\bf A}$. So, the equation of motion of the
guided atoms takes the form $\dot{\bf p}=\mu{\bf
v}\times\left(\nabla\times{\bf A}\right)=-2\eta \mu{\bf
v}\times({{\bf k}\times \dot{\bf k}})/{k^{2}}$, namely,
\begin{equation}
\dot{\bf k}+2\eta {\bf k}\times\left(\frac{{\bf k}\times\dot{\bf
k}}{k^{2}}\right)=0.
\end{equation}
Now we can determine the parameter $\eta$ in both the Lagrangian
and the above equation. It is apparently seen that when $\eta$ is
taken to be $1/2$, the obtained $\dot{\bf k}+{\bf k}\times({{\bf
k}\times\dot{\bf k}})/{k^{2}}=0$ is an identity. In what follows,
we take $\eta=1/2$.

Because of $\nabla\cdot{\bf A}=0$, the above phenomenological
Hamiltonian can be rewritten as $H={\bf P}^{2}/2\mu-{\bf
A}\cdot{\bf P}+(\mu/2){\bf A}^{2}$, where the term $-{\bf
A}\cdot{\bf P}$ is
\begin{equation}
-{\bf A}\cdot{\bf P}=\frac{1}{2}\left(\frac{{\bf k}\times \dot{\bf
k}}{k^{2}}\right)\cdot {\bf J},     \label{AP}
\end{equation}
where ${\bf J}={\bf r}\times{\bf P}$, which is the total angular
momentum of both guided atoms and guiding potential fields. It
should be noted that the above analysis is treated merely inside
the classical Newtonian framework. If one utilizes the equivalence
principle, the effective Hamiltonian that describes such a
phenomenological interaction is two times that of the obtained
result (\ref{AP})\cite{Mashhoon,Zhu,Shenprb}. Thus, the total
effective Hamiltonian is
\begin{equation}
H_{\rm eff}(t)=\left[\frac{{\bf k}(t)\times \dot{\bf
k}(t)}{k^{2}}\right]\cdot {\bf J},
\end{equation}
and consequently the equation governing the time evolution of the
atomic matter wave inside the noncoplanarly curved atom fibre is
of the form
\begin{equation}
H_{\rm eff}(t)|m, {\bf k}(t)\rangle=i\hbar\frac{\partial}{\partial
t}|m, {\bf k}(t)\rangle.     \label{eqA3}
\end{equation}
The solution to the above equation has been presented in
Ref.\cite{Shen2}. In the following, we simply mention the
principal result of Ref.\cite{Shen2}. In order to solve Eq.
(\ref{eqA3}), the atom wave vector can be rewritten in the
spherical polar coordinate system, {\it i.e.}, ${\bf
k}(t)=k(\sin\theta\cos\varphi, \sin\theta\sin\varphi,
\cos\theta)$, where both $\theta$ and $\varphi$ are the
time-dependent functions. It is assumed that the initial wave
vector is ${\bf k}(0)=(0, 0, k)$, {\it i.e.}, the initial polar
angle $\theta(0)=0$. According to the Lewis-Riesenfeld invariant
formulation\cite{Lewis,Shenchen}, the wavefunction in Eq.
(\ref{eqA3}) can be written in the form $ |m, {\bf
k}(t)\rangle=\exp\left[\frac{1}{i}\phi_{m}(t)\right]V(t)|m\rangle$,
where $|m\rangle$ is the initial state satisfying the eigenvalue
equation $J_{3}|m\rangle=m\hbar|m\rangle$, and the expression for
the phase is
$\phi_{m}(t)=m\int^{t}_{0}\dot{\varphi}(t')\left[1-\cos\theta
(t')\right]{\rm d}t'$. The time-dependent operator $V(t)$ takes
the form $ V(t)=\exp\left[\beta(t) J_{+}-\beta^{\ast}(t)
J_{-}\right] $ with $\beta=-(\theta/2)\exp(-i\varphi)$ and
$\beta^{\ast}=-(\theta/2)\exp(i\varphi)$. Here the operators
$J_{\pm}=\left(J_{1}\pm iJ_{2}\right)/\hbar$.

In the above, we obtained the atomic matter-wave motional state of
one-dimensional longitudinal propagation in the curved atom fibre.

\section{Atomic matter waves in the presence of a guiding potential field}
A number of ways has been suggested for building atomic
matter-wave guides in experiments. These include the methods by
using electric (electrostatic) forces\cite{Hau,Seka,Den}, magnetic
forces\cite{Den2,Lau,Hinds}, and light-induced
forces\cite{Ito,yIN,Torii}. Here, we will briefly consider the
scheme of electric (electrostatic) and magnetic forces, and then
investigate in details the time evolution of the interacting
system (including the atomic internal levels, light fields and
atomic centre of mass motion) in the scheme of light-induced
forces.

\subsection{Electrostatic field as a guiding field}
The techniques of electric-magnetic traps and guides can be
utilized to realize the monomode atom wave guides and are
integrated into a mesoscopic ``atom chip'' technology. The
interaction Hamiltonian of the electrostatic fields and neutral
but polarizable atoms (polarizability $\alpha$) is given $H_{\rm
e-a}=-\alpha E^{2}/2$\cite{Kruger}, which leads to the propagation
of atomic matter wave along the one-dimensional longitudinal
direction inside the atom wave guide. Such a motion results from
the mechanism that the atomic centre of mass will be drawn towards
the larger electric fields since the interaction between the
induced electric dipole and the electric field is always
attractive. To realize a stable trapping configuration on atom
chips, one can use a combination of the electric and magnetic
interactions. The full Hamilton of such a three-dimensional
confinement is $H=\mu_{\rm B}g_{\rm F}m_{\rm F}B-\alpha
E^{2}/2$\cite{Kruger}, where $\mu_{\rm B}, g_{\rm F}, m_{\rm F},
B$ denote the Bohr magneton, the Land\'{e} factor, the magnetic
quantum number, and the magnetic field modulus, respectively.
Here, the role of the magnetic interaction is to realize a
magnetic wire guide (side guide), in which atoms are trapped in a
potential tube along a line parallel to a straight current
carrying wire\cite{Luo}. Recently, Luo {\it et al.} reported an
experiments of trapping and manipulating neutral atoms with
electrostatic fields\cite{Kruger}. More recently, this group
reported on the implementation and experimental test of a key
element for the controlled manipulation of matter waves on the
atom chip: an omnidirectional atom fibre\cite{Luo}. As to the
theoretical treatment for the scheme of electrostatic and magnetic
forces, readers may be referred to
Refs.\cite{Stenholm1,Stenholm2,Stenholm3,Folman5,Rohwedder,Stenholm4}.

\subsection{Light-induced potential as a guiding potential}

The total Hamiltonian of the atomic matter wave is $H\left({\bf
r},t\right) ={p^{2}}/{2\mu}+H_{a}\left({\bf r},t\right)$. Here
$H_{a}\left({\bf r},t\right)$ denotes the Hamiltonian of the
atomic internal levels coupled to a guiding potential field, and
has the following form
\begin{eqnarray}
H_{a}\left({\bf r},t\right)&=&\frac{1}{2}\hbar \Delta\left(| {\rm
e}\rangle\langle {\rm e}|-|{\rm g}\rangle \langle {\rm
g}|\right)-{\bf p}_{\rm eg}\cdot {\bf E}\left({\bf
r},t\right)|{\rm g}\rangle\langle {\rm e}|       \nonumber \\
& &   -{\bf p}_{\rm ge}\cdot {\bf E}\left({\bf r},t\right) |{\rm
e}\rangle\langle {\rm g}|,
\end{eqnarray}
where $|{\rm e}\rangle$ and $|{\rm g}\rangle$ denote the ground
and excited states of the two-level atom, respectively, and the
frequency detuning $\Delta=\omega _{\rm e}-\omega$. Here $\omega
_{\rm e}$ and $\omega$ are the atomic transition frequency and the
light-field mode frequency, respectively. In the representation of
the two-state base vectors $\left\{|{\rm e}\rangle, |{\rm
g}\rangle\right\}$, the matrix form of $H_{a}\left({\bf
r},t\right)$ reads
\begin{equation}
H_{a}\left({\bf r},t\right) =\left(
\begin{array}{cc}
{\frac{1}{2}\hbar \Delta}  & {V\left({\bf r},t\right)}  \\
{V^{\ast}\left({\bf r},t\right)}  & {-\frac{1}{2}\hbar \Delta}
\end{array}
\right),        \label{Hamiltonian}
\end{equation}
where $V\left({\bf r},t\right)=-{\bf p}_{\rm ge}\cdot{\bf
E}\left({\bf r},t\right)$ and $V^{\ast}\left({\bf
r},t\right)=-{\bf p}_{\rm eg}\cdot{\bf E}\left({\bf r},t\right) $.

Here we only consider the ultracold atoms, whose centre of mass
motion cannot easily drive the transitions between the atomic
internal levels. If the external field varies spatially rather
adiabatically, then the Born-Oppenheimer approximation is
applicable to this problem, and the wavefunction, $\Psi
_{\eta}\left({\bf r},t\right)$, of the atomic system can be
separable: it can be rewritten as the product of the
centre-of-mass wavefunction $\Phi _{\eta}\left({\bf r},t\right)$
and the atomic internal-level eigenstates $|\psi \left({\bf
r},t;\eta\right) \rangle$, {\it i.e.}, $\Psi _{\eta}\left({\bf
r},t\right) =\Phi _{\eta}\left({\bf r},t\right) |\psi \left({\bf
r},t;\eta\right) \rangle $. Because of the validity of the
Born-Oppenheimer approximation, the derivatives of $|\psi
\left({\bf r},t;\eta\right) \rangle$ with respect to the spatial
coordinates can be ignored in the following calculation. Thus, the
motional state of the atomic centre of mass and the wavefunction
of atomic internal levels agree with the following time-dependent
Schr\"{o}dinger equations
\begin{equation}
\left[\frac{p^{2}}{2\mu}+U\left({\bf r},t\right) \right] \Phi
_{\eta}\left({\bf r},t\right) =i\hbar\frac{\partial}{\partial
t}\Phi _{\eta}\left({\bf r},t\right)    \label{centremass}
\end{equation}
and
\begin{equation}
\left[H_{a}\left({\bf r},t\right) -U\left({\bf r},t\right) \right]
| \psi \left({\bf r},t;\eta \right)\rangle
=i\hbar\frac{\partial}{\partial t} |\psi \left({\bf r},t;\eta
\right)\rangle,   \label{internal}
\end{equation}
respectively. If one substitutes the relation $ | \psi \left({\bf
r},t;\eta \right)\rangle =\exp \left[
({i}/{\hbar})\int_{0}^{t}U\left({\bf r},t'\right){\rm d}t'\right]
| \psi \left({\bf r},t;\eta \right)\rangle _{U} $ into Eq.
(\ref{internal}), then one can obtain
\begin{equation}
H_{a}\left({\bf r},t\right) | \psi \left({\bf r},t;\eta
\right)\rangle _{U}=i\hbar\frac{\partial}{\partial t}| \psi
\left({\bf r},t;\eta \right) \rangle_{U}.      \label{Schrodinger}
\end{equation}

In the following, we will solve the time-dependent Schr\"{o}dinger
equation (\ref{Schrodinger}).

\subsection{Time evolution in the presence of light fields}

The study of topics on the time evolution in atom optics is of
interest. For example, recently, several proposals for atomic
matter-wave interferometers have been made in the context of
splitting and combining a microtrap ground state with a
time-dependent potential\cite{Folman5,Hinds2,Reichel2}. Here, we
also consider the behavior of atomic matter wave in the presence
of a time-dependent guiding potential field.

For convenience, in the following we will temporarily not mention
the variable ${\bf r}$ in $V$. The Hamiltonian (\ref{Hamiltonian})
can be written in the form
$H_{a}=v_{1}\sigma_{1}+v_{2}\sigma_{2}+(\hbar\Delta/2)\sigma_{3}$,
where $\sigma_{i}$'s denote the Pauli matrices, and
$V=v_{1}-iv_{2}$, $V^{\ast}=v_{1}+iv_{2}$. Thus, $H_{a}(t)$ can be
rewritten as
\begin{eqnarray}
H_{a}(t)&=&\hbar\Omega(t)\{\frac{1}{2}\sin
\theta(t)\exp\left[-i\phi(t)\right]S_{+}       \nonumber \\
& &  +\frac{1}{2}\sin
\theta(t)\exp\left[i\phi(t)\right]S_{-}+\cos\theta(t)S_{3} \},
\label{Hamiltonian2}
\end{eqnarray}
where the instantaneous energy eigenvalue
$\hbar\Omega(t)=2\sqrt{\hbar^{2}\Delta^{2}/4+V^{\ast}(t)V(t)}$,
$\theta=\arccos\left(\Delta/\Omega\right)$ and
$\phi=\arctan(v_{2}/v_{1})$. The operators $S_{i}=(1/2)\sigma_{i}$
($i=1,2,3$) and $S_{\pm}=S_{1}\pm iS_{2}$, and agree with the
commutation relations $\left[S_{+},S_{-}\right]=2S_{3}$ and
$\left[S_{3}, S_{\pm}\right]=\pm S_{\pm}$. According to the
Lewis-Riesenfeld invariant theory\cite{Lewis}, the
Lewis-Riesenfeld invariant that has time-independent eigenvalues
obeys the following Liouville-von Neumann equation\cite{Lewis}
\begin{equation}
\frac{\partial}{\partial t}I(t)+\frac{1}{i\hbar}\left[I(t),
H(t)\right]=0.    \label{Liouville}
\end{equation}
The eigenvalue equation of the invariant $I(t)$ is
$I(t)|t;\eta\rangle=\eta |t;\eta\rangle$, where the eigenvalue
$\eta$ is time-independent, {\it i.e.}, $\partial\eta/\partial
t=0$. As the Hamiltonian is constructed in terms of the three
operators $S_{3}$ and $S_{\pm}$, the invariant $I(t)$ in Eq.
(\ref{Liouville}) should also take the form expressed by these
operators. Thus, we have
\begin{eqnarray}
I(t)&=&\frac{1}{2}\sin
\varsigma(t)\exp\left[-i\zeta(t)\right]S_{+}
 \nonumber \\
& &  +\frac{1}{2}\sin
\varsigma(t)\exp\left[i\zeta(t)\right]S_{-}+\cos\varsigma(t)
S_{3}.   \label{invariant}
\end{eqnarray}
The time-dependent parameters $\varsigma(t)$ and $\zeta(t)$ can be
determined by Eq. (\ref{Liouville}). Insertion of Eq.
(\ref{invariant}) into Eq. (\ref{Liouville}) yields the following
set of auxiliary equations
\begin{equation}
\left\{
\begin{array}{ll}
&   \exp\left(-i\zeta\right)\left(\dot{\varsigma}\cos
\varsigma-i\dot{\zeta}\sin \zeta\right)-i\Omega
[\exp(-i\phi)\cos\varsigma\sin\theta
 \nonumber \\
& \qquad \qquad     \qquad   \qquad  \qquad  \quad    -\exp(-i\zeta)\sin\varsigma\cos\theta]=0,            \\
&   \dot{\varsigma}+\Omega \sin\theta \sin(\zeta-\phi)=0,
\end{array}
\right. \label{auxiliary}
\end{equation}
which can determine the parameters $\varsigma(t)$ and $\zeta(t)$
of the Lewis-Riesenfeld invariant (\ref{invariant}).

It follows from the Lewis-Riesenfeld invariant theory that the
particular solution $| \psi \left({\bf r},t;\eta \right)
\rangle_{U}$ of the time-dependent Schr\"{o}dinger equation
(\ref{Schrodinger}) is different from the eigenstate $\left|t;
\eta\right\rangle$ of the invariant $I(t)$ only by a
time-dependent c-number factor
$\exp\left[\frac{1}{i}\varphi_{\eta}(t)\right]$, namely, the
general solutions of Eq. (\ref{Schrodinger}) can be written in the
form
\begin{equation}
|\Psi(t)\rangle=\sum_{\eta}C_{\eta}\exp\left[\frac{1}{i\hbar}\varphi_{\eta}(t)\right]\left|t;
\eta\right\rangle,           \label{phase}
\end{equation}
where $C_{\eta}$ stands for some certain time-independent
coefficients, and the time-dependent phases in (\ref{phase}) are
given by
\begin{equation}
\varphi_{\eta}(t)=\int_{0}^{t}\langle t';
\eta|\left[H(t')-i\hbar\frac{\partial}{\partial t'}\right]|t';
\eta\rangle {\rm d}t'.      \label{phase2}
\end{equation}

In view of the above discussion, one can see that if the
eigenstate $|t; \eta\rangle $ is obtained, then the time-dependent
Schr\"{o}dinger equation (\ref{Schrodinger}) is solved by the
Lewis-Riesenfeld invariant theory\cite{Lewis}. To solve the
eigenvalue equation $I(t)|t;\eta\rangle=\eta |t;\eta\rangle$ of
the invariant, one can utilize a time-dependent unitary
transformation $V(t)$, which leads to\cite{Gao}
\begin{equation}
\left[V^{\dagger}(t)I(t)V(t)\right]V^{\dagger}(t) |t;
\eta\rangle=\eta V^{\dagger}(t)|t; \eta\rangle.
\label{unitaryinvariant}
\end{equation}
It is believed that by choosing the appropriate parameters in
$V(t)$, one can obtain $V^{\dagger}(t)I(t)V(t)=S_{3}$ and
$|\eta\rangle=V^{\dagger}(t)|t; \eta\rangle$, and thus Eq.
(\ref{unitaryinvariant}) will be rewritten as a very simple form
$S_{3}|\eta\rangle=\eta |\eta\rangle$, which is easily solved.
Such a time-dependent unitary transformation operator takes the
form\cite{Shenchen}
\begin{equation}
V(t)=\exp\left\{
\left[-\frac{\varsigma(t)}{2}e^{-i\zeta(t)}\right]S_{+}-\left[-\frac{\varsigma(t)}{2}e^{i\zeta(t)}\right]S_{-}\right\}.
\end{equation}
By using such a time-dependent unitary transformation, the
time-dependent phase (\ref{phase2}) can be rewritten as
\begin{equation}
\varphi_{\eta}(t)=\int_{0}^{t}\langle
\eta|\left[V^{\dagger}(t')H(t')V(t')-i\hbar
V^{\dagger}(t')\frac{\partial V(t')}{\partial
t'}\right]|\eta\rangle {\rm d}t'.
\end{equation}
Further calculation shows that
$\varphi_{\eta}(t)=\varphi_{\eta}^{({\rm
d})}(t)+\varphi_{\eta}^{({\rm g})}(t)$, where the dynamical phase
is
\begin{eqnarray}
\varphi_{\eta}^{({\rm d})}(t)&=&\eta
\int_{0}^{t}\hbar\Omega(t')\{\cos\varsigma(t')\cos\theta(t')
    \nonumber \\
& &
+\sin\varsigma(t')\sin\theta(t')\cos\left[\zeta(t')-\phi(t')\right]\}{\rm
d}t',        \label{dynamical}
\end{eqnarray}
and the geometric phase is
\begin{equation}
\varphi_{\eta}^{({\rm g})}(t)=\eta
\int_{0}^{t}\hbar\dot{\zeta}(t')\left[1-\cos\varsigma
(t')\right]{\rm d}t'.    \label{geometric}
\end{equation}

As the eigenvalues of $S_{3}$ are $\pm 1/2$ corresponding to the
eigenstates $\left|\eta=\pm \frac{1}{2}\right\rangle$, the
eigenvalues of the invariant $I(t)$ are $\eta=\pm 1/2$. Thus, the
eigenfunctions of the invariant are as follows $ \left|t; \eta=\pm
\frac{1}{2}\right\rangle=V(t)\left|\eta=\pm
\frac{1}{2}\right\rangle$, which can be rewritten as
\begin{equation}
\left\{
\begin{array}{ll}
&  \left|t;
\eta=+\frac{1}{2}\right\rangle=\cos\frac{\varsigma(t)}{2}\left(
{\begin{array}{*{20}c}
   {1}  \\
   {0}  \\
\end{array}} \right)+e^{i\zeta(t)}\sin
\frac{\varsigma(t)}{2}\left( {\begin{array}{*{20}c}
   {0}  \\
   {1}  \\
\end{array}} \right),   \\
&  \left|t;
\eta=-\frac{1}{2}\right\rangle=\cos\frac{\varsigma(t)}{2}\left(
{\begin{array}{*{20}c}
   {0}  \\
   {1}  \\
\end{array}} \right)-e^{-i\zeta(t)}\sin
\frac{\varsigma(t)}{2}\left( {\begin{array}{*{20}c}
   {1}  \\
   {0}  \\
\end{array}} \right),
\end{array}
\right. \label{invarianteigen}
\end{equation}
where $\left|\eta=\pm\frac{1}{2}\right\rangle$ have been expressed
by the two-dimensional column matrices, {\it i.e.},
\begin{equation}
\left|\eta=+\frac{1}{2}\right\rangle=\left( {\begin{array}{*{20}c}
   {1}  \\
   {0}  \\
\end{array}} \right),   \qquad    \left|\eta=-\frac{1}{2}\right\rangle=\left( {\begin{array}{*{20}c}
   {0}  \\
   {1}  \\
\end{array}} \right).
\end{equation}

\subsection{Discussions}

In what follows, we will discuss the adiabatic case. It is
apparent that when the Hamiltonian (\ref{Hamiltonian2}) is
time-independent, {\it i.e.}, the parameters $\theta={\rm const.}$
and $\phi={\rm const.}$, according to the auxiliary equation
(\ref{auxiliary}), the solution of the parameters in the invariant
(\ref{invariant}) is $\varsigma=\theta$, $\zeta=\phi$. For the
adiabatic case, $\varsigma(t)\simeq \theta(t)$ and $\zeta(t)\simeq
\phi(t)$ can be viewed as the approximate solution of the
auxiliary equation (\ref{auxiliary}). So, in the adiabatic
approximation, the eigenbasis (\ref{invarianteigen}) of the
invariant $I(t)$ will be reduced to the following form
\begin{equation}
\left\{
\begin{array}{ll}
&  \left|t;
\eta=+\frac{1}{2}\right\rangle=\cos\frac{\theta}{2}\left(
{\begin{array}{*{20}c}
   {1}  \\
   {0}  \\
\end{array}} \right)+e^{i\phi}\sin\frac{\theta}{2}\left(
{\begin{array}{*{20}c}
   {0}  \\
   {1}  \\
\end{array}} \right),   \\
&   \left|t;
\eta=-\frac{1}{2}\right\rangle=\cos\frac{\theta}{2}\left(
{\begin{array}{*{20}c}
   {0}  \\
   {1}  \\
\end{array}} \right)-e^{-i\phi}\sin\frac{\theta}{2}\left(
{\begin{array}{*{20}c}
   {1}  \\
   {0}  \\
\end{array}} \right).
\end{array}
\right. \label{invarianteigen2}
\end{equation}

The general solution of the time-dependent Schr\"{o}dinger
equation (\ref{Schrodinger}) is
\begin{equation}
|\Psi(t)\rangle=c_{+}(t)\left|t;
\eta=+\frac{1}{2}\right\rangle+c_{-}(t)\left|t;
\eta=-\frac{1}{2}\right\rangle,
\end{equation}
where the time-dependent coefficients are defined as $
c_{\pm}(t)\propto \exp\left[-i\varphi_{\eta=\pm
{1}/{2}}(t)/\hbar\right]$.

According to the two expressions (\ref{dynamical}) and
(\ref{geometric}), the phase in the time-dependent coefficients
$c_{\pm}(t)$ is expressed by $\varphi_{\eta}(t)\simeq \eta
\int_{0}^{t}\hbar\Omega(t'){\rm d}t'$. In the meanwhile, the
invariant $I(t)$ is also reduced to the Hamiltonian, {\it i.e.},
$I\rightarrow H/\Omega$. This, therefore, means that in the
adiabatic approximation, the eigenvalue equation of the invariant
acts as the instantaneous eigenvalue equation of the Hamiltonian.
In the nonadiabatic case, however, the instantaneous eigenvalue
equation of the Hamiltonian is no longer valid, and it should be
replaced with the eigenvalue equation of the Lewis-Riesenfeld
invariant.

Now we consider the atomic centre of mass motion in the presence
of external fields. For the stationary (or adiabatic) case, it
follows from Eqs. (\ref{centremass}), (\ref{internal}) and
(\ref{Schrodinger}) that the eigenvalue equation of atomic centre
of mass motion is
\begin{equation}
\left[\frac{p^{2}}{2\mu}\pm
\sqrt{\frac{\hbar^{2}\Delta^{2}}{4}+V^{\ast}({\bf r})V({\bf
r})}\right] \Phi _{\pm}\left({\bf r}\right) =E\Phi
_{\pm}\left({\bf r}\right). \label{centremass2}
\end{equation}

In principle, we presented a theoretical treatment for the time
evolution of the atom-light interacting system.

\section{Atomic matter wave in a weakly guiding field}

\subsection{Atomic matter wave in a quantized light field}

The full Hamiltonian of the two-level atomic ensemble interacting
with a photon field is $ H={p^{2}}/{2\mu}+H_{\rm A-F}$, where
$H_{\rm A-F}$ denotes the total Hamiltonian that includes the free
Hamiltonians of the atomic internal levels, the external photon
field as well as their interaction Hamiltonian. $H_{\rm A-F}$ is
written in the form
\begin{eqnarray}
 H_{\rm A-F}&=& \frac{1}{2}\hbar \omega _{\rm e}\left(| {\rm
e}\rangle \langle {\rm e}| -| {\rm g}\rangle \langle {\rm
g}|\right) +\hbar [g\left({\bf r},t\right) a| {\rm e}\rangle
\langle {\rm g}|             \nonumber \\
& &   +g^{\ast}\left({\bf r},t\right) a^{\dagger}| {\rm g}\rangle
\langle {\rm e}|] +\hbar \omega a^{\dagger}a,     \label{AF}
\end{eqnarray}
where $a$ and $a^{\dagger}$ stand for the annihilation and
creation operators of the photons, respectively. The coupling
coefficient in the atom-field Hamiltonian (\ref{AF}) is $ {g}({\bf
r},t)=-{E_{0}}{\bf d}\cdot{\bf u}^{\ast}({\bf r},t)/{\hbar}$,
where $E_{0}=\sqrt{\hbar\omega/2\epsilon _{0}{\mathcal V}}$ stands
for the electric field strength of monophoton, and ${\bf d}$ and
${\bf u}$ denote the electric-dipole transition matrix element and
the electric-field normal mode, respectively. In the expression
for $E_{0}$, the quantization volume $ {\mathcal V}=\int{\bf
u}^{\ast}\cdot{\bf u}{\rm d}{\bf r}$. According to the
Born-Oppenheimer approximation, the wavefunction, $\Psi_{\eta,
m}({\bf r},t)$, of the system under consideration can be
separable, {\it i.e.},
\begin{equation}
\Psi_{\eta, m}({\bf r},t)=\Phi _{\eta,m}\left({\bf r},t\right)
V_{\omega}(t) | \psi _{m}\left({\bf r},t;\eta \right) \rangle,
\label{separable}
\end{equation}
where $\Phi _{\eta,m}\left({\bf r},t\right)$ and $| \psi
_{m}\left({\bf r},t;\eta \right) \rangle$ denote the motional
state of the atomic centre of mass and the wavefunction of atomic
internal levels, respectively. The time-dependent unitary
transformation $V_{\omega}(t)$ in (\ref{separable}) takes the form
\begin{equation}
V_{\omega}(t) =\exp \left[\frac{1}{i} \omega \left (\frac{| {\rm
e}\rangle \langle {\rm e}| -| {\rm g}\rangle \langle {\rm
g}|}{2}+a^{\dagger}a\right) t\right].
\end{equation}
It follows that the motional state $\Phi _{\eta,m}\left({\bf
r},t\right)$ of the atomic centre of mass and the wavefunction $|
\psi _{m}\left({\bf r},t;\eta \right) \rangle$ of atomic internal
levels satisfy the following two time-dependent Schr\"{o}dinger
equations
\begin{equation}
\left[\frac{p^{2}}{2\mu}+U_{\eta ,m}({\bf r},t)\right] \Phi _{\eta
,m}\left({\bf r},t\right) =i\hbar \frac{\partial}{\partial t}\Phi
_{\eta ,m}\left({\bf r},t\right)
\end{equation}
and
\begin{equation}
\left[H_{A}({\bf r},t)-U_{\eta ,m}({\bf r},t)\right]
|\psi_{m}\left({\bf r},t;\eta \right) \rangle =i\hbar
\frac{\partial}{\partial t} |\psi _{m}\left({\bf r},t;\eta \right)
\rangle,            \label{HAU}
\end{equation}
respectively. Here the Hamiltonian $H_{\rm A}({\bf r},t)$ is
\begin{eqnarray}
H_{\rm A}({\bf r},t)&=& \frac{1}{2}\hbar \Delta\left(| {\rm
e}\rangle \langle {\rm e}|-|{\rm g}\rangle \langle {\rm g}|\right)
 \nonumber \\
& &   +\hbar \left[{g}\left({\bf r},t\right) a| {\rm e}\rangle
\langle {\rm g}|+{g}^{\ast}\left({\bf r},t\right) a^{\dagger}|
{\rm g}\rangle \langle {\rm e}|\right].       \label{internals}
\end{eqnarray}

We should simplify Eq. (\ref{HAU}). By inserting the relation $ |
\psi _{m}\left({\bf r},t;\eta \right) \rangle =\exp \left[
({i}/{\hbar})\int_{0}^{t}U_{\eta ,m}({\bf r},t'){\rm d}t'\right]
|\psi_{m}\left({\bf r},t; \eta \right) \rangle_{U} $ into Eq.
(\ref{HAU}), one can yield
\begin{equation}
H_{\rm A}({\bf r},t)|\psi _{m}\left({\bf r},t; \eta \right)
\rangle _{U}=i\hbar \frac{\partial}{\partial t}| \psi
_{m}\left({\bf r},t;\eta \right)\rangle_{U}.       \label{HAU2}
\end{equation}

In the representation of the two-state base vectors $\left\{|{\rm
e}\rangle, |{\rm g}\rangle\right\}$, the operator $|{\rm e}\rangle
\langle {\rm e}|-|{\rm g}\rangle \langle {\rm g}|$ can be
rewritten as a matrix form, {\it i.e.}, $|{\rm e}\rangle \langle
{\rm e}|-|{\rm g}\rangle \langle {\rm g}|=\sigma _{3} $. Thus, one
can obtain the commutation relation $\left[a|{\rm e}\rangle
\langle {\rm g}|, a^{\dagger}|{\rm g}\rangle \langle {\rm
e}|\right] =N'\sigma _{3}$, where the operator $N'$ is
\begin{equation}
N'=\left(\begin{array}{cc}
{aa^{\dagger}} & {0} \\
{0} & {a^{\dagger}a}
\end{array}
\right).
\end{equation}
The eigenvalue equation of $N'$ is as follows
\begin{equation}
N'\left( {\begin{array}{*{20}c}
   {| m\rangle}  \\
   {| m+1\rangle}  \\
\end{array}} \right)=(m+1) \left( {\begin{array}{*{20}c}
   {| m\rangle}  \\
   {| m+1\rangle}  \\
\end{array}} \right).
\label{eigeneq}
\end{equation}
Note that the operator $N'$ commutes with all the generators in
the Hamiltonian (\ref{internals}). For this reason, the
eigenstates of $N'$ have an essential significance in treating Eq.
(\ref{HAU}) or (\ref{HAU2}), namely, the solutions of Eq.
(\ref{HAU2}) can be obtained in the sub-Hilbert-space
corresponding to the particular eigenvalue of the operator $N'$.

If we set the operators $S_{+}=(m+1)^{-\frac{1}{2}}a| {\rm
e}\rangle \langle {\rm g}|$,
$S_{-}=(m+1)^{-\frac{1}{2}}a^{\dagger}| {\rm g}\rangle \langle
{\rm e}|$ and $S_{3}=({1}/{2})\sigma _{3}$, we can obtain the
following commutation relations $[S_{+}, S_{-}]=2S_{3}$ and
$[S_{3}, S_{\pm}]=\pm S_{\pm}$. Thus, the Hamiltonian
(\ref{internals}) of the atomic internal levels (coupled to a
radiation field) can be rewritten as
\begin{equation}
H_{\rm A}({\bf r},t)=\hbar \Delta S_{3}+\hbar \sqrt{m+1}\left[
{g}\left({\bf r},t\right) S_{+}+{g}^{\ast}\left({\bf r},t\right)
S_{-}\right].         \label{new}
\end{equation}

\subsection{Adiabatic eigenstates}
It is apparent that the Hamiltonian (\ref{new}) has a same form as
(\ref{Hamiltonian2}). Therefore, one can easily obtain the
time-evolution solution of Eq. (\ref{HAU2}) by using the method
presented in the preceding section. So, here we will not further
discuss this problem. Instead, we will consider the stationary (or
adiabatic) case for this quantum mechanical system. Calculation
shows that the eigenstates of the atom-field Hamiltonian
(\ref{AF}) are as follows
\begin{equation}
\left\{
\begin{array}{ll}
&    |{\bf r};m,+\rangle=\cos\frac{\theta_{m}}{2}|{\rm e}, m\rangle-\sin \frac{\theta_{m}}{2}|{\rm g}, m+1\rangle,                               \\
&     |{\bf r};m,-\rangle=\cos\frac{\theta_{m}}{2}|{\rm g},
m+1\rangle+\sin \frac{\theta_{m}}{2}|{\rm e}, m\rangle
\end{array}
\right. \label{eigenstates}
\end{equation}
corresponding to the energy eigenvalues
\begin{equation}
U_{\pm, m}({\bf r})=\left(m+\frac{1}{2}\right)\hbar\omega \pm
\sqrt{\frac{\hbar^{2}\Delta^{2}}{4}+\left(m+1\right)\left[g({\bf
r})\right]^{2}},
\end{equation}
respectively. The parameter angle $\theta_{m}$ in the eigenstates
(\ref{eigenstates}) is defined as $ \theta_{m}({\bf
r})=\arctan\left[{-2\sqrt{m+1}g({\bf r})}/{\Delta}\right]$, where
$g({\bf r})$ is assumed to be a real function.

In the meanwhile, the motional state of the atomic centre of mass
satisfies
\begin{equation}
\left[\frac{p^{2}}{2\mu}+U_{\pm ,m}({\bf r})\right] \Phi _{\pm,
m}\left({\bf r}\right) =E\Phi _{\pm, m}\left({\bf r}\right).
\label{lightinduced}
\end{equation}
This means that the atom in the presence of a light field will
undergo a light-induced force, which can be used to guide atomic
matter waves.

\subsection{Atomic matter-wave bandgap structure in a weakly guiding field}

External fields (light, electrostatic field and magnetic field)
can cool, guide and trap atoms. Apart from these roles which
external fields play, there may be a new effect of atomic matter
wave arising from the interaction between atomic internal levels
and external optical fields, {\it i.e.}, the existence of the {\it
atomic matter-wave bandgap structure} in a spatially-periodic
guiding field or a cavity field. As the spatially-periodic
potential (or force) acting upon the atoms is provided by the
optical fields, such a medium can also be referred to as ``{\it
optical lattice bandgap medium}". One of the properties of this
optical lattice bandgap medium (atomic matter-wave bandgap medium)
is such that some atoms with certain momentum can propagate
through the potential region, and other atoms which have momentum
within the atomic band gap will, however, be recoiled and
therefore cannot propagate in the atomic matter-wave bandgap
medium. As the ``lattice constant'' of the spatially periodic
potential (optical lattice) is the wavelength of the optical
field, in order to form such an {\it optical lattice bandgap
structure}, the atomic momentum should be of about $10^{-27}$
Kg$\cdot$m/s. Thus, such a kind of atomic media (optical lattice
bandgap media) can be
realized below $T=10^{-5}$ K. 

During the last decades, a kind of materials referred to as
photonic crystals, which is patterned with a periodicity in
dielectric constant and can therefore create a range of forbidden
frequencies called a photonic band gap, focus considerable
attention of many researchers\cite{Yablonovitch,John}. Likewise,
we believe that it is also of physical interest to investigate its
atomic counterpart, particularly in the field of Bose-Einstein
condensation (BEC). The photonic crystal is, however, a
classically electromagnetic material, since the behavior of light
can be governed by the classical Maxwellian equations. Such an
atomic (optical lattice) bandgap structure medium is a purely
quantum optical material. Since it will offer the possibility of
molding the flow of atoms in the guiding potential fields, this
optical lattice bandgap structure may enable the atomic waveguide
technique to be achieved more effectively in the technology of
atom chips.

Additionally, it has been suggested that the atoms and molecules
can be trapped inside a single-wall carbon nanotube. Recently, a
new molecular conformation of carbon, {\it i.e.}, the
supramolecular assembly comprising C$_{60}$ molecules inside
single-wall carbon nanotubes was synthesized by using transmission
electron microscopy\cite{Smith} and vapor phase
methods\cite{Smith2}. As the encapsulated C$_{60}$ molecules with
a neighbor distance of 1 nm are in the carbon nanotubes, such a
molecular assembly is referred to as the nanoscopic carbon
peapod\cite{Smith}. The potential application of the carbon peapod
is such that it can be utilized as a memory device capable of
storing a bit of information by the voltage-driven shuttling of
interior C$_{60}$ between the two endcaps of the single-wall
carbon nanotubes\cite{Smith2}. We believe that in addition to the
method of controlling the C$_{60}$ molecules trapped in the
nanoscopic carbon peapod by using the moderate voltage, it can be
manipulated by the optical fields in the carbon nanotube. Such
optical fields itself has also been trapped in the carbon peapod
because the electric permittivity inside the carbon nanotube is of
spatial periodicity. Thus, the carbon peapod can be considered a
one-dimensional photonic crystal. In the next section, we will
propose a theoretical mechanism for the coherent control of
C$_{60}$ molecules trapped in the nanoscopic carbon peapod by both
the external controlling light and the interior optical fields
trapped in the carbon nanotube.

\section{Coherent control of three-level atomic matter wave by an
external controlling field}

In this section, a scheme to manipulate atomic matter waves inside
the wave guide by using an external controlling field is
suggested. If, for example, such guided atoms have a system of
three-level configuration, the full Hamiltonian of which is
$H={p^{2}}/{2\mu}+H_{\rm A-F}\left({\bf r}\right)$, where the
atom-field Hamiltonian $H_{\rm A-F}\left({\bf r}\right)$ reads
\begin{eqnarray}
H_{\rm A-F}\left({\bf r}\right)&=& \hbar\omega^{\rm e}|{\rm
e}\rangle\langle{\rm e}|
+\hbar\omega^{{\rm g}_{1}}|{\rm {\rm g}_{1}}\rangle\langle{\rm {\rm g}_{1}}|
+\hbar\omega^{{\rm g}_{2}}|{\rm {\rm g}_{2}}\rangle\langle{\rm {\rm g}_{2}}|                \nonumber \\
& &
+\hbar\omega_{1}a_{1}^{\dagger}a_{1}+\hbar\omega_{2}a_{2}^{\dagger}a_{2}                  \nonumber \\
& & +\hbar g_{1}\left({\bf r}\right)\left(a_{1}|{\rm
e}\rangle\langle{\rm {\rm g}_{1}}|+a_{1}^{\dagger}|{\rm {\rm
g}_{1}}\rangle\langle{\rm e}|\right)              \nonumber \\
& & +\hbar g_{2}\left({\bf r}\right)\left(a_{2}|{\rm
e}\rangle\langle{\rm {\rm g}_{2}}|+a_{2}^{\dagger}|{\rm {\rm
g}_{2}}\rangle\langle{\rm e}|\right).         \label{Hamil}
\end{eqnarray}
Here we assume that the three-level atom has one upper (excited)
and two lower (ground) states, which are denoted by $|{\rm
e}\rangle$, $|{\rm {\rm g}_{1}}\rangle$ and $|{\rm {\rm
g}_{2}}\rangle$, respectively. The corresponding energy
eigenvalues of these three states are $\hbar\omega^{\rm e}$,
$\hbar\omega^{{\rm g}_{1}}$ and $\hbar\omega^{{\rm g}_{2}}$,
respectively. $\omega_{1}$ and $\omega_{2}$ in (\ref{Hamil})
denote the mode frequencies of the two photon fields coupled to
the $|{\rm {\rm g}_{1}}\rangle$-$|{\rm e}\rangle$ and $|{\rm {\rm
g}_{2}}\rangle$-$|{\rm e}\rangle$ transitions, the coupling
coefficients of which are $g_{1}$ and $g_{2}$, respectively. By
using the Born-Oppenheimer approximation, the separable
wavefunction $\Psi_{\eta, n_{1}, n_{2}}\left({\bf r}\right)$ can
be written as $\Psi_{\eta, n_{1}, n_{2}}\left({\bf
r}\right)=\Phi_{\eta, n_{1}, n_{2}}\left({\bf r}\right)|{\bf r};
n_{1}, n_{2}, \eta\rangle$, where $\Phi_{\eta, n_{1},
n_{2}}\left({\bf r}\right)$ and $|{\bf r}; n_{1}, n_{2},
\eta\rangle$ stand for the motional state of atomic centre of mass
and the eigenstate of the interaction between atomic internal
levels and two radiation fields, the latter of which agrees with
the eigenvalue equation $H_{\rm A-F}\left({\bf r}\right)|{\bf r};
n_{1}, n_{2}, \eta\rangle={\mathcal E}_{\eta, n_{1},
n_{2}}\left({\bf r}\right)|{\bf r}; n_{1}, n_{2}, \eta\rangle$
with $\eta=0, \pm$. Its three eigenstates are given as follows
\begin{eqnarray}
|{\bf r}; n_{1}, n_{2}, 0\rangle &=& {\mathcal
N}_{0}(g_{2}\sqrt{n_{2}+1}|{\rm
 g}_{1},n_{1}+1,n_{2}\rangle                   \nonumber \\
& &    -g_{1}\sqrt{n_{1}+1}|{\rm g}_{2},n_{1},n_{2}+1\rangle),                \nonumber \\
|{\bf r}; n_{1}, n_{2}, \pm\rangle &=& {\mathcal N}_{\pm}[|{\rm
 e},n_{1},n_{2}\rangle     \nonumber \\
 & & +k_{\pm}(g_{1}\sqrt{n_{1}+1}|{\rm
 g}_{1},n_{1}+1,n_{2}\rangle                \nonumber \\
& & +g_{2}\sqrt{n_{2}+1}|{\rm g}_{2},n_{1},n_{2}+1\rangle)],
\label{threeeigenstates}
\end{eqnarray}
where the parameter
\begin{equation}
k_{\pm}=\frac{1}{2}\left[\frac{{\mathcal N}_{0}^{2}{\mathcal
E}_{0, n_{1}, n_{2}}}{\hbar}\pm \sqrt{\left(\frac{{\mathcal
N}_{0}^{2}{\mathcal E}_{0, n_{1},
n_{2}}}{\hbar}\right)^{2}+4{\mathcal N}_{0}^{2}}\right].
\end{equation}
The normalized coefficients in (\ref{threeeigenstates}) are
${\mathcal N}_{0}=1/\sqrt{g_{1}^{2}(n_{1}+1)+g_{2}^{2}(n_{2}+1)}$,
${\mathcal N}_{\pm}=1/\sqrt{1+k^{2}_{\pm}/{\mathcal N}_{0}^{2}}$.
The corresponding energy eigenvalues of the three eigenstates
(\ref{threeeigenstates}) are given by
\begin{eqnarray}
{\mathcal E}_{0, n_{1}, n_{2}}&=& \hbar\left(\omega^{\rm e}+n_{1}\omega_{1}+n_{2}\omega_{2}\right),                \nonumber \\
{\mathcal E}_{\pm, n_{1}, n_{2}}&=& \frac{1}{2}\left({\mathcal
E}_{0, n_{1}, n_{2}} \pm \sqrt{{\mathcal E}_{0, n_{1},
n_{2}}^{2}+\frac{4\hbar^{2}}{{\mathcal N}_{0}^{2}}}\right).
\end{eqnarray}

Thus the state $\Phi_{\eta, n_{1}, n_{2}}\left({\bf r}\right)$ of
the atomic centre of mass motion satisfies
\begin{equation}
\left[\frac{p^{2}}{2\mu}+{\mathcal E}_{\eta, n_{1}, n_{2}}({\bf
r})\right]\Phi_{\eta, n_{1}, n_{2}}=E\Phi_{\eta, n_{1},
n_{2}}\left({\bf r}\right).       \label{lightinduced2}
\end{equation}

It should be noted that the presented formulation can be applied
to the investigation of the controllable manipulation of
three-level atomic matter-wave bandgap structure by the external
controlling light. It follows from Eq. (\ref{lightinduced2}) that
the atomic matter-wave bandgap structure caused by one of the
optical fields, say, the field coupled to the $|{\rm {\rm
g}_{1}}\rangle$-$|{\rm e}\rangle$ level pair, can be controllably
manipulated by another optical fields ({\it i.e.}, the optical
field coupled to the $|{\rm {\rm g}_{2}}\rangle$-$|{\rm e}\rangle$
pair).

\section{Concluding remarks}
The motion of atoms inside the guides in both temporal and spatial
domains were discussed in the present paper. We considered four
topics: (i) a phenomenological model to describe the wave
propagation of atoms in the {\it curved} atom guides; (ii) the
motion of atomic matter wave in the presence of both classical and
quantized guiding fields; (iii) the existence of the atomic
matter-wave bandgap (optical lattice bandgap) structure in a
cavity field or a spatially periodic guiding field; (iv) the
controllable manipulation of three-level atomic matter wave in a
wave guide by an external controlling light. The paper provided a
theoretical formulation for treating the time evolutional process
of the atomic matter wave by the time-dependent guiding potential.
One of the most remarkable concepts in this paper is the atomic
matter-wave bandgap structure. We pointed out that such an atomic
bandgap medium can be coherently controlled by external fields,
which may have some potential applications in information
technology, {\it e.g.}, the technique of information storage.
\\ \\
\textbf{Acknowledgements}  This work was supported partially by
both the National Natural Science Foundation of China (Project
Nos. $90101024$ and $60378037$) and the National Basic Research
Program (973) of China (Project No. 2004CB719805).

\end{document}